\documentclass[aps,prd,reprint,superscriptaddress,nofootinbib]{revtex4-1}

\usepackage{amsmath,amssymb,graphicx,bm}
\usepackage[colorlinks=true, linkcolor=blue, citecolor=blue, urlcolor=blue]{hyperref}
\usepackage{float}
\usepackage{xcolor,verbatim}
\newcommand{\be}{\begin{equation}}
\newcommand{\ee}{\end{equation}}

\newcommand{\bea}{\begin{eqnarray}}
\newcommand{\eea}{\end{eqnarray}}
\newcommand{\nn}{\nonumber}
\newcommand{\sss}{\scriptscriptstyle}

\begin{document}
\rightline{FERMILAB-PUB-26-0176-T}
\title{Neutrinos as Dark Matter}

\author{James M. Cline}
\email{jcline@physics.mcgill.ca}
\affiliation{McGill University Department of Physics $\&$ Trottier Space Institute,
3600 Rue University, Montréal, QC, H3A 2T8, Canada}
\author{Gonzalo Herrera}
\email{gonzaloh@mit.edu}
\affiliation{Physics Department and Kavli Institute for Astrophysics and Space Research, Massachusetts Institute of Technology, Cambridge, MA 02139, USA}
\affiliation{Harvard University, Department of Physics and Laboratory for Particle Physics and Cosmology, Cambridge, MA 02138, USA}
\author{Jean-Samuel Roux}
\email{jean-samuel.roux@mail.mcgill.ca}
\affiliation{McGill University Department of Physics $\&$ Trottier Space Institute,
3600 Rue University, Montréal, QC, H3A 2T8, Canada}

\begin{abstract}
Active neutrinos in standard cosmology were ruled out as a dark matter candidate in the 1980's. The reason is twofold: they are too light to account for the observed energy density of dark matter in the Universe, and their relativistic nature would spoil structure formation. In this note we suggest that an enhanced density of cold Standard Model active neutrinos today could behave effectively as dark matter, avoiding constraints from recombination and structure formation. Such an enhancement could be produced, for instance, by late-time decays of a light scalar field that is not in thermal equilibrium with the plasma. This mechanism is testable through the detection of the Cosmic Neutrino Background (C$\nu$B), which could have an average cosmological energy density a factor of $\sim 100-200$ times larger than expected in $\Lambda$CDM. The postulated light neutrinophilic scalar field may be observable, with Yukawa couplings in the range $y \sim 5 \times 10^{-16}-10^{-12}$. A scenario preferred by structure formation constraints is that the scalar is a Majoron, and the neutrinos have an inverted mass hierarchy.
\end{abstract}

\maketitle

\emph{\textbf{Introduction.}}
Active neutrinos, despite being abundant, neutral and stable, cannot account for the observed dark matter abundance in the Universe within the standard cosmological framework. Active neutrinos decouple from the thermal plasma at temperatures of order $T \sim 1$ MeV, while still relativistic. Their comoving number density is preserved after decoupling and is related to the photon number density as \cite{Baumann:2022mni}
\begin{equation}
\frac{n_\nu}{n_\gamma} = \frac{3}{11}.
\end{equation}
The present-day number density per flavor is approximately $n_\nu \simeq 112\, \mathrm{cm}^{-3}$ \cite{Weinberg:1962zza, Kolb:1990vq}. If neutrinos have masses $m_\nu \lesssim 0.4$ eV \cite{KATRIN:2024cdt}, their total energy density today is then
\begin{equation}
\Omega_\nu h^2 = \frac{\sum_i m_{\nu_i}}{94\, \rm eV} \lesssim 0.013,
\end{equation}
which is nearly an order of magnitude below the observed dark matter density $\Omega_{\rm DM} h^2 \simeq 0.12$ \cite{Planck:2018vyg}. Thus, active relic neutrinos are too light to constitute the bulk of dark matter (DM).

Moreover, standard relic neutrinos are relativistic at decoupling and remain so until late times. This causes them to free-stream out of gravitational potential wells, erasing small-scale structure. This is tightly constrained by the cosmic microwave background (CMB) and the matter power spectrum, since it suppresses structure formation below scales
\begin{equation}
\lambda_{\rm FS} \sim \frac{1}{H_0} \left( \frac{T_\nu}{m_\nu} \right) \sim 10 \, \mathrm{Mpc} \left(\frac{0.1 \mathrm{eV}}{m_\nu}\right)\,,
\end{equation}
with Hubble rate $H_0 \simeq 70 \mathrm{~km} / \mathrm{s} / \mathrm{Mpc}$  and present day $\nu$ temperature $T_\nu \simeq 1.95 \mathrm{~K} \simeq 1.7 \times 10^{-4} \mathrm{eV}$.

Hence, Standard Model relic neutrinos behave as hot DM, requiring them to be a small fraction of $\Omega_{\rm DM}$.  
These arguments assume the standard phase-space distribution of relic neutrinos in the early Universe, which might be changed by new physics. A viable model of 
active neutrino DM, having
$\Omega_{\nu}^{\rm total} \simeq \Omega_{\rm DM} \simeq 0.26$ and consistency with experimental constraints, must
satisfy several conditions:

\begin{enumerate}

\item The production mechanism must not spoil Big Bang Nucleosynthesis (BBN), which is sensitive to any additional energy density present at
$T\sim 0.1\,{\rm MeV}$, since it modifies the Hubble expansion rate and the weak interaction freeze-out of neutrons. An enhanced neutrino population prior to BBN is must have sufficiently small energy density and impact on $\nu n\leftrightarrow p e$ rates. The effective number of extra $\nu$ species is constrained by $\Delta N_{\rm eff}\lesssim 0.3$. Either the enhanced $\nu$ population is produced after BBN, or with energies below $T \sim 0.1$ MeV and out of equilibrium, if produced earlier.

\item The new production mechanism injects neutrinos with sufficiently low momenta, or early enough such that they redshift to become cold by matter-radiation equality ($T_{\rm eq} \sim 0.8$ eV).
Alternatively, they may become the dark matter after recombination, while the parent particles behave as dark matter in the CMB.

\item The enhanced relic neutrino population must remain cosmologically cold and stable until today, and it must be compatible with bounds from neutrino capture in laboratory experiments \cite{KATRIN:2022kkv} and searches for cosmic-ray boosted C$\nu$B flux \cite{Herrera:2026pzj}. It should
and must respect the Pauli exclusion principle on different physical scales. This restricts the fraction of the dark matter abundance that active neutrinos are allowed to constitute at small scales.
\end{enumerate}

If neutrinos are the bulk of DM at present, and have  masses in the range $m_\nu \sim 0.05 - 0.4$ eV, the C$\nu$B density is enhanced by a factor of 
\begin{equation}
n_\nu^{\rm DM} \simeq 28-226\, n_\nu\,.
\end{equation}
We will argue that the exclusion of active neutrinos as dark matter is not generic, but instead relies crucially on assumptions about their thermal origin and phase-space distribution, as specified  by the above conditions. In the following, we present a minimal possible realization.
\bigskip

\begin{figure}[t]
    \centering
    \includegraphics[width=0.5\textwidth]{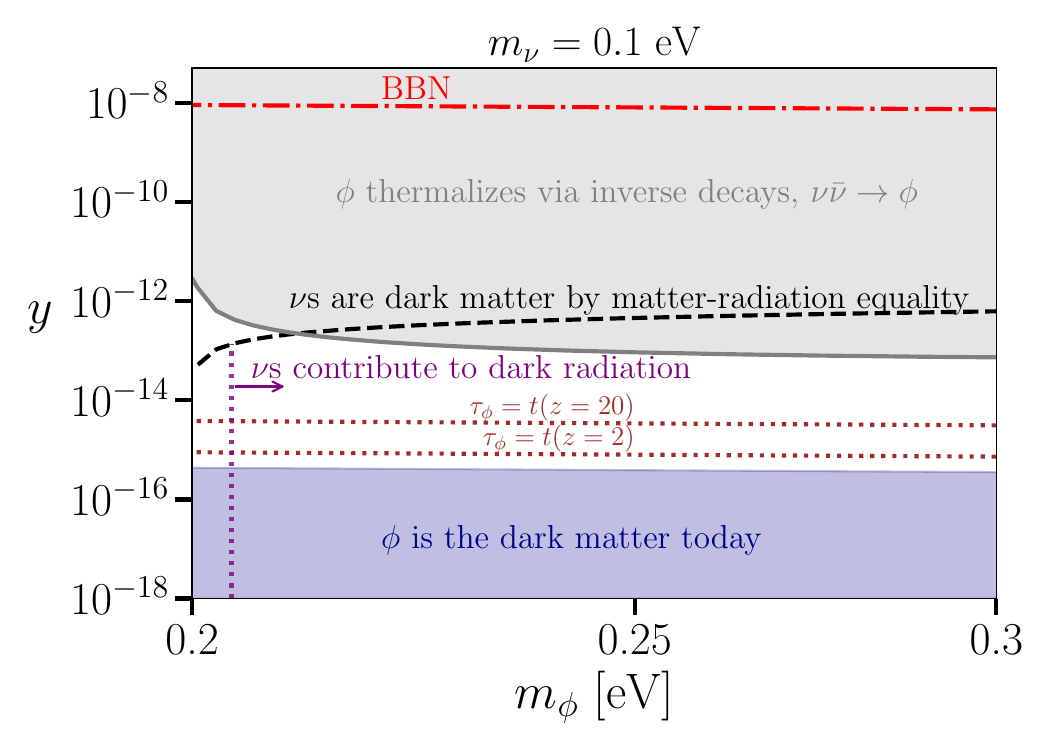} 
    \caption{Yukawa coupling of light scalar $\phi$ to neutrinos $\nu$ versus $m_{\phi}$, with  $m_{\nu}=0.1$ eV representing the heaviest active neutrino mass.  Shaded regions are ruled out as labeled on the figure.
    Horizontal dotted lines indicate
    decays occuring at redshifts 2 and 20.}
    
    \label{fig:parameter_space}
\end{figure}

\emph{\textbf{Enhanced cold neutrino population from non-thermal scalar decays.}} 
A minimal extension of the Standard Model that can fulfill our requirements is a light (pseudo)scalar field $\phi$ that couples exclusively to neutrinos via a Yukawa interaction. It may be a low-energy effective description of a more complete model, for instance Majorons \cite{Gelmini:1980re, Chikashige:1980ui,Brune:2018sab}.  A singlet Majoron $\phi$ is a pseudo-Nambu-Goldstone boson appearing as the phase of a complex scalar field
$\Phi = \rho e^{i\phi/f}$, whose real part gets a vacuum expectation value (VEV) $\langle\rho\rangle = f/\sqrt{2}$ that spontaneously breaks lepton number. The  interaction of $\phi$ with neutrinos takes the form
\begin{equation}
\label{yukawa-lag}
\mathcal{L} \supset y_i\,\phi\,\bar{\nu}_i\,\gamma_5\,\nu_i,
\qquad
y_i \equiv \frac{m_{\nu_i}}{f},
\end{equation}
where $f$ is the lepton-number breaking scale and the coupling is diagonal in the neutrino mass ($m_{\nu_i}$) basis. This interaction can be written in a gauge-invariant form as the dimension-six operator
$\phi(\bar L H)(i\gamma_5)(HL)/M^2$, which after electroweak symmetry breaking reproduces
Eq.~(\ref{yukawa-lag}).
The pseudoscalar $\phi$ is assumed to never thermalize with the plasma; its cosmological abundance instead arises  non-thermally. Here we will focus on the misalignment mechanism.

We assume that lepton number is spontaneously broken before inflation,
so that $\phi$ is homogenized over the observable Universe and any topological defects are inflated away \cite{Abbott:1982af, Preskill:1982cy}. The radial mode of the complex scalar is assumed to be heavy ($m_{\rho} \gg H_I$) and settled at its minimum, leaving only the angular $\phi$ degree of freedom relevant for late-time dynamics.

Suppose $\phi$ is displaced from its minimum during inflation to an initial value $\phi_0$. The scalar field will oscillate once $H\lesssim m_\phi$, at a temperature $T_{\rm osc}^4= 90\, m_\phi^2 M_p^2/(\pi^2 g_*)$. At this point, its abundance $Y_\phi$ becomes fixed,  
\be \label{eq:yphi}
Y_\phi = \frac{\rho_\phi}{m_\phi s} = \frac{45}{4\pi^2 } \frac{m_\phi \phi_0^2}{g_{*s}T_{\rm osc}^3} \sim 0.01\, \frac{m_\phi \phi_0^2}{T_{\rm{osc}}^3}\,,
\ee 
taking $g_{*\sss S}\sim 100$. Assuming the scalar field completely decays, the neutrino abundance will be enhanced by $Y_\nu\sim 2 Y_\phi$. For $m_\nu \sim 0.1$ eV, active neutrinos make up all of DM if $Y_\nu \sim 4$ \cite{Cline:2018fuq}. Using Eq.\ (\ref{eq:yphi}), this implies
\be \label{eq:misalign}
\phi_0\sim 1.26\times 10^{12} \ {\rm GeV}\ \left(\frac{m_\phi}{0.2\rm \ eV}\right)^{1/4}\,.
\ee 
The required initial amplitude is safely below the Planck scale, which ensures that $\phi$ will oscillate before its vacuum energy dominates.

The initial displacement of $\phi$ is determined by  the stochastic misalignment mechanism \cite{Graham:2018jyp, Cline:2024wja}. Assuming inflation lasts for a sufficiently long time, the average displacement within a Hubble patch is $\langle \phi_0^2\rangle=3H_I^4/(8 \pi^2 m_\phi^2)$, where $H_I^2=\Lambda_I^4/(3M_p^2)$ is the Hubble parameter during inflation. Assuming reheating is efficient, $\Lambda_I^4 = (\pi^2/30) g_* T_{\rm RH}^4$, one finds
\be 
T_{\rm RH} \sim 3.3\times 10 ^9\ {\rm GeV}\ \left(\frac{m_\phi}{0.2\ \rm{eV}}\right)^{5/16}.
\ee 
This determines $\Lambda_I\sim8\times 10^{9}$ GeV, hence $H_I\sim15$ GeV\footnote{This value of  $H_I$ is orders of magnitude below the bound from isocurvature; see the Supplementary Material \ref{sec:isocurvature_constraints}.}.

\bigskip

\emph{\textbf{Decay of scalar to neutrinos.}}
The scalar particle decays to neutrinos with rate
\begin{equation}\label{eq:decay_rate}
\Gamma(\phi \rightarrow \nu \bar{\nu})=
\frac{y^2 m_\phi}{8 \pi}
\left(1-\frac{4 m_\nu^2}{m_\phi^2}\right)^{1/2}.
\end{equation}
If the scalar population was not thermalized and is therefore cold, decays prior to BBN do not disrupt light-element abundances, for $m_\phi\ll 1$\,MeV. Their contribution to $N_{\rm eff}$ is further bounded to $\Delta N_{\rm eff} \lesssim 0.3$.
For reference, the values of the Yukawa coupling corresponding to decays at BBN ($t \sim 10^3\,{\rm s}$) is indicated as a dotted-dashed red line in Fig. \ref{fig:parameter_space}.

If the neutrinos are the dark matter by matter-radiation equality, then they must be nonrelativistic by $t_{\rm eq } \simeq 10^{12} \mathrm{~s}$. This implies a velocity at $t_{\rm eq}$ of 
\begin{equation}
v_{\nu, \rm eq}=\frac{p_\nu\left(t_{\mathrm{eq}}\right)}{m_\nu}=\frac{m_\phi}{2 m_\nu}\left(\frac{a_{\rm decay }}{a_{\mathrm{eq}}}\right) \ll 1 .
\end{equation}
During radiation domination, $a(t) \propto t^{1 / 2}$, so this translates into a maximum allowed decay time
\begin{equation}
\tau_\phi^{\max }\left(m_\phi\right)={t_{\mathrm{eq}}}\left(\frac{2 m_\nu}{m_\phi}\right)^2 .
\end{equation}
This condition is shown as a dashed black line in Fig.\ \ref{fig:parameter_space}.

However, $\phi$ can play the role of dark matter during the initial stage of structure formation.
It must then
decay within the age of the Universe for the neutrinos to become the DM. This sets a lower limit on the coupling, 
\begin{equation}
y \gtrsim \sqrt{\frac{8 \pi \hbar}{m_\phi t_U}} \gtrsim 4.39 \times 10^{-16}\left(\frac{0.2 \, \mathrm{eV}}{m_\phi}\right)^{1 / 2}
\end{equation}
with $t_U \sim 4.35 \times 10^{17} \mathrm{~s}$. It excludes the region in blue in Fig. \ref{fig:parameter_space}.

To keep the $\phi$ particles out of equilibrium, the rate of inverse decays $\nu \bar{\nu} \rightarrow \phi$
must be smaller than the Hubble expansion rate at temperatures  $T\gtrsim m_\phi$, hence $\Gamma_\phi \lesssim H(T = m_\phi)$,
giving the bound
\begin{equation}
y^2 \lesssim
{8 \pi \, 1.66 \sqrt{g_*(m_\phi)}}
{m_\phi\over M_{\mathrm{Pl}}}\left(1 - \frac{4 m_\nu^2}{m_\phi^2}\right)^{-1/2}\,,
\end{equation}
or numerically,
\begin{equation}
y \lesssim
3.5 \times 10^{-14}
\left( \frac{m_\phi}{0.2~\mathrm{eV}} \right)^{1/2}
\left[ 1 - \left( \frac{0.2~\mathrm{eV}}{m_\phi} \right)^2 \right]^{-1/4},
\end{equation}

which excludes the grey region in In Fig.~\ref{fig:parameter_space}. Such small values 
also make freeze-in \cite{Hall:2009bx} inefficient for producing $\phi$ particles.

\bigskip

{\emph{\textbf{Constraints from structure formation.}} CMB data constrain the fraction of post-recombination
conversion of dark matter into invisible radiation to be at
most a few percent, $F_{\rm max} = 0.038$~\cite{Poulin:2016nat,Bringmann:2018jpr}.
For two-body decays at rest, the effective radiation fraction
scales as
\begin{equation}
F_{\rm rad,eff} = F \sum_i {\rm BR}_i \, v_i^2 \,,
\label{eq:Fradeff}
\end{equation}
with final-state velocity
$v_i = \sqrt{1 - 4m_{\nu_i}^2/m_\phi^2}$,
$F$ denoting the decayed fraction of dark matter,
and ${\rm BR}_i$ the branching ratio into eigenstate~$i$.
In the Majoron realization with diagonal couplings
$y_i = m_{\nu_i}/f$, the partial widths scale as
$\Gamma_i \propto m_{\nu_i}^2\, v_i$, correlating the
branching ratios with the velocities.

For normal ordering ($m_3 \gg m_2 > m_1$),
the heaviest eigenstate dominates the total width.
Taking $F \simeq 1$, the two lightest states are produced
nearly relativistically ($v_{1,2} \simeq 1$), contributing
significantly to $F_{\rm rad,eff}$.
Minimizing over $m_\phi$ yields
\begin{equation}
F_{\rm rad,eff}^{\rm min} \simeq 0.16 \,,
\label{eq:Fmin_NO}
\end{equation}
well above $F_{\rm max} = 0.038$.
Normal ordering therefore cannot satisfy the CMB bound
if $F \simeq 1$, though it remains viable if the decays occur
predominantly before recombination, or if only a
subdominant fraction $F < 1$ decays at late times.

For inverted ordering
($m_1 \simeq m_2 \equiv m_h \gg m_3$),
choosing $m_\phi \simeq 2\,m_h$ produces two near-threshold
channels with small velocities, while the lightest eigenstate
$\nu_3$ carries negligible branching ratio for
$m_{\nu_3} \ll m_h$.
The post-recombination bound is naturally satisfied,
requiring only the mild condition
\begin{equation}
m_{\nu_3} \lesssim 3.7 \times 10^{-3}\,{\rm eV}
\left(\frac{m_h}{0.05\,{\rm eV}}\right) ,
\label{eq:mnu3_bound}
\end{equation}
which enforces near-threshold decays and motivates the
purple line in Fig.~\ref{fig:parameter_space}.
A detailed derivation, including the dependence on
$m_\phi/(2\,m_{\nu,\rm heavy})$, is given in the
Supplementary Material~\ref{sec:fradiation}.

\begin{figure*}[t!]
    \centerline{
    \includegraphics[width=0.55\textwidth]{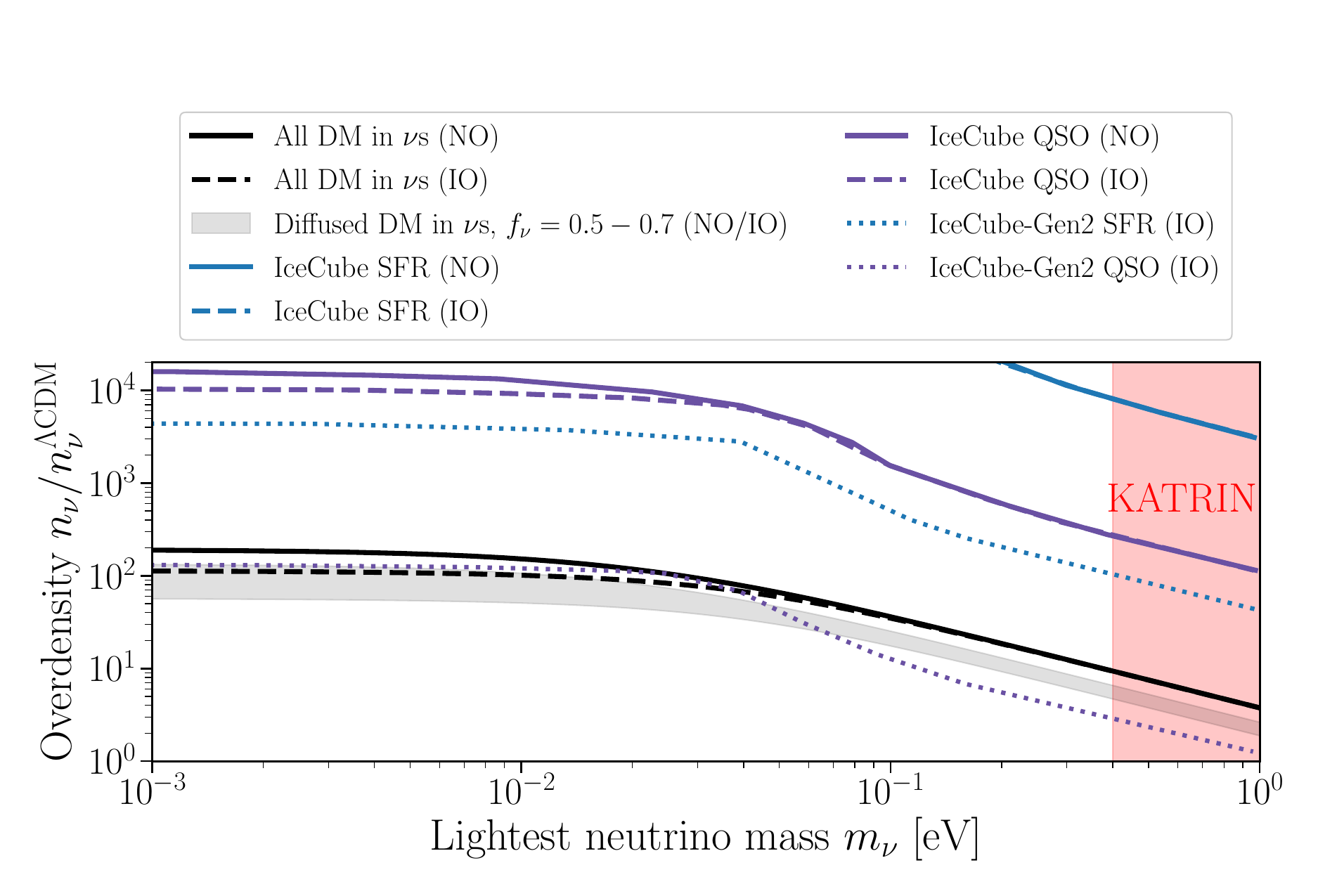} 
     \raisebox{0.9em}{\includegraphics[width=0.45\textwidth]{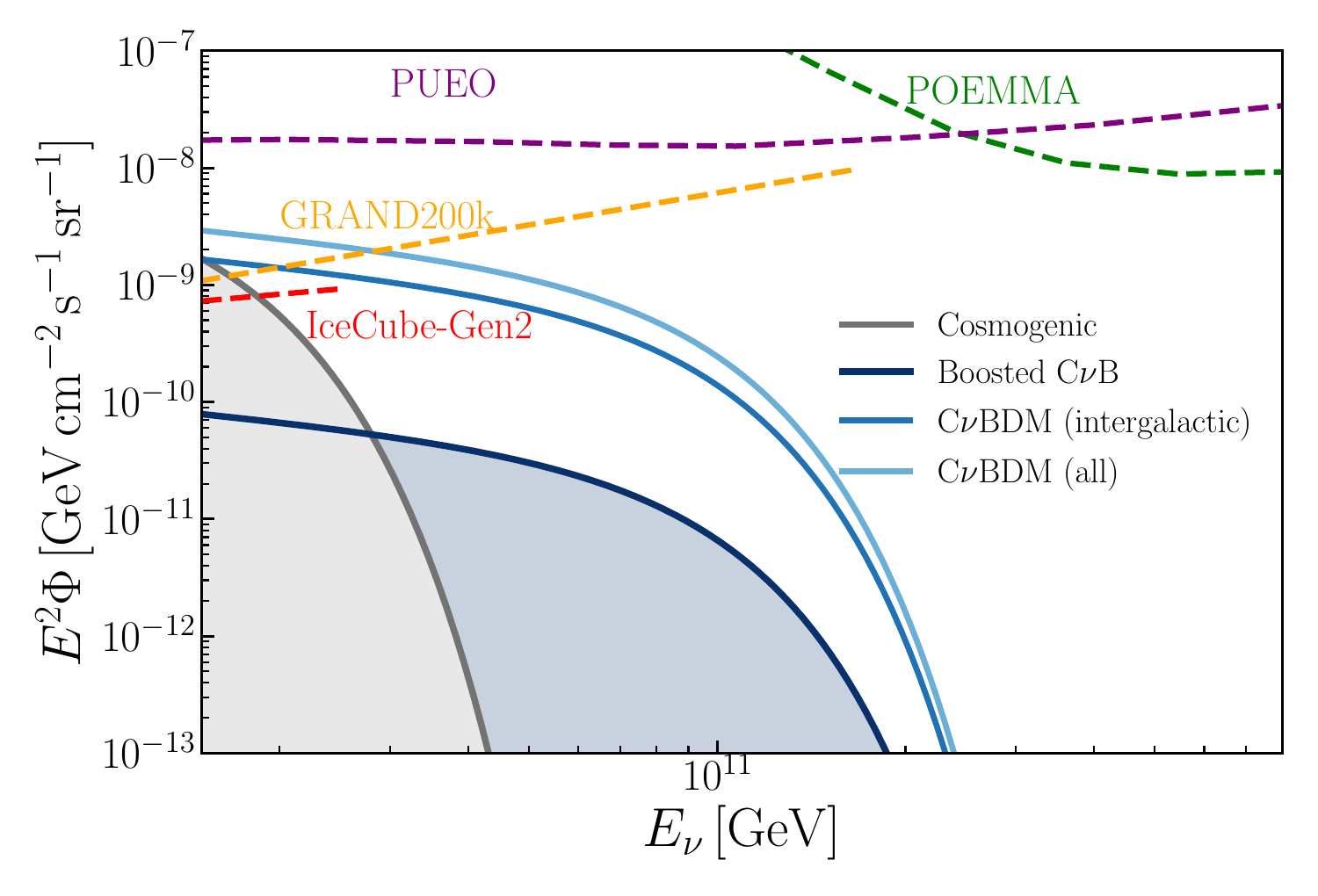}} }    
    \caption{Left: Predicted enhancement relative to the $\Lambda$CDM cosmic neutrino background, as a function of the lightest neutrino mass. Solid black line: overdensity if active neutrinos are all of the dark matter. Shaded grey band: a fraction of the DM is in neutrinos on diffuse cosmological scales, while $\phi$ is rest of DM at small scales. Red region: excluded by KATRIN \cite{KATRIN:2024cdt}. Colored curves:  limits from IceCube on the Diffuse Boosted Cosmic Neutrino Background \cite{Herrera:2026pzj}, for normal (NO) and inverted (IO) neutrino mass ordering, assuming star-formation–rate (SFR) and quasar (QSO) source evolution models. Dotted curves: projected sensitivity of IceCube-Gen2. Right: Illustration of the high-energy cut-off of cosmogenic neutrinos versus cosmic-ray boosted cosmic neutrino background, for an assumed maximum cosmic-ray energy of $E^{\rm max}_{\rm CR}=10^{12}$GeV. The resulting fluxes for an enhanced cosmic neutrino background as dark matter as predicted in this work are also shown, for $m_{\nu}=0.1$ eV. A different maximum cosmic-ray energy would shift the cut-off energy of the cosmogenic and boosted cosmic neutrino background fluxes, but the cross-over is fixed by particle physics, given by the different inelasticities of both processes. Estimated projected sensitivities from PUEO, POEMMA, GRAND200k and IceCube-Gen2 are shown for comparison \cite{Ackermann:2022rqc}.} 
    \label{fig:limits}
\end{figure*}

\bigskip
\emph{\textbf{Phase space constraints.}}
Free fermions in a gravitational potential obey the Tremaine-Gunn bound on the maximum coarse-grained phase-space density \cite{Tremaine:1979we}, coming from the Pauli exclusion principle. 
For an escape velocity $v_{\rm esc}$ this implies a maximum neutrino density
\be
\rho_\nu^{\max}\simeq \frac{m_\nu^4 v_{\rm esc}^3}{3\pi^2},
\label{eq:pauli_main}
\ee
and hence a maximal allowed dark matter decayed fraction into cold, gravitationally bound neutrinos
\begin{equation}
F_{\rm Pauli}^{\rm max}
\simeq
1.6\times10^{-3}\,
\left(\frac{m_\nu}{0.1\,\mathrm{eV}}\right)^4
\left(\frac{v_{\rm esc}}{500\,\mathrm{km\,s^{-1}}}\right)^3
\left(\frac{\bar\rho_{\rm DM}}{\rho_{\rm DM}}\right)\,,
\end{equation}
with $\bar\rho_{\rm DM}\equiv \Omega_{\rm DM}\rho_c \simeq 1.3\times10^{-6}\,{\rm GeV\,cm^{-3}}$ and $\rho_{\rm DM}$ denoting the average DM density of the system. Thus, in dense galactic halos only a small fraction of $\phi$ can convert to free neutrinos, whereas on large diffuse scales Pauli blocking is weak, allowing neutrinos to constitute most of the cosmological dark matter.

Furthermore, if the decayed neutrinos constitute the bulk of the dark matter on such scales, then they must be sufficiently slow by the time small-scale structures form. Dwarf galaxies impose a stringent restriction on the allowed neutrino velocities as a function of redshift \begin{equation}
v_\nu(z_f) \;\lesssim\; v_{\rm esc}^{\rm dwarf},
\end{equation}
where $v_{\rm esc}^{\rm dwarf} \lesssim 80~\mathrm{km\,s^{-1}}$ is a
representative escape velocity for dwarf halos. This bound can be translated into a bound on the scalar mass, which yields a fine tuning at least larger than $1\%$. A detailed numerical discussion is provided in the Supplemental Material, Section \ref{sec:pauli}. In some models, this fine-tuning can be relaxed, further preventing decays in dense structures while allowing them in larger scales, see Section \ref{sec:screening}.

\bigskip
\emph{\textbf{Experimental tests.}} 
Our proposed scenario could be tested in several ways. The most striking prediction  is an enhanced C$\nu$B relative to the standard expectation in $\Lambda$CDM. If active neutrinos were all of the dark matter today, their 
number density would be $\Omega_\nu=\Omega_{\rm DM}$, implying
\begin{equation}
\begin{aligned}
n_{\nu \rm DM}
&\simeq
\frac{\Omega_{\rm DM}}{\Omega_\nu}\,n_\nu
\simeq
\left(28-226\right)\,n_\nu
\\
&\simeq
(3.1\times10^{3}-2.5\times10^{4})~\mathrm{cm^{-3}} \,,
\label{nrelest}
\end{aligned}
\end{equation}
for neutrino masses in the range $m_\nu\simeq0.05$-$0.4$~eV, where
$n_\nu\simeq112~\mathrm{cm^{-3}}$ is the standard relic density per flavor, and $m_{\nu}<0.4$ from the KATRIN experiment \cite{KATRIN:2024cdt}.
However, accounting for Pauli phase-space constraints on small scales, neutrinos can be only a fraction
$F_{\nu}=0.5-0.7$ of the total dark matter on diffuse scales. This reduces the estimate (\ref{nrelest}) to
\begin{equation}
\begin{aligned}
n_{\nu \rm DM}^{\rm diff}
&\simeq
F_\nu\,n_{\nu \rm DM}
\simeq
(14-158)\,n_\nu
\\
&\simeq
(1.6\times10^{3}-1.8\times10^{4})~\mathrm{cm^{-3}} .
\end{aligned}
\end{equation}

Direct detection searches exist, by neutrino capture on $\beta$-decaying nuclei with KATRIN or PTOLEMY \cite{Weinberg:1962zza,KATRIN:2022kkv, PTOLEMY:2019hkd}. KATRIN currently constraints a local overdensity of the C$\nu$B at $10^{11}$ times the cosmological average \cite{KATRIN:2022kkv}. However, given the Tremaine-Gunn bound, the enhancement of the C$\nu$B within the Milky Way from the kinematically-allowed decays of the scalar would be of order $\sim 1\%$, which is smaller than other possible enhancements, for instance from gravitational clustering \cite{Ringwald:2004np}.

More promising is a possible detection of the diffuse boosted cosmic neutrino background induced by cosmic ray-neutrino scatterings over cosmological scales \cite{Ciscar-Monsalvatje:2024tvm,DeMarchi:2024zer,Herrera:2024upj,Zhang:2025rqh,Herrera:2026pzj}. IceCube currently constrains a C$\nu$B overdensity at the level of $10^{3}-10^{5}$ for $m_{\nu} \gtrsim 0.1$ eV, with the exact limit depending on the ultra-high energy cosmic ray flux evolution with redshift. IceCube-Gen2 could potentially test overdensities at the level of 10 to 100 times the cosmological average, falling within the theoretical predictions of our model; see Fig.\ \ref{fig:limits}. The induced boosted C$\nu$B flux also has a larger endpoint energy than cosmogenic neutrinos, due to the higher inelasticity of the scattering process \cite{Formaggio:2012cpf, Weigel:2024gzh}. More details are provided in the Supplementary Material \ref{sec:testability_appendix}.

Indirect tests exist by detection of the sub-eV scale neutrinophilic scalar. Current constraints on the Yukawa coupling are however orders of magnitude weaker ($y \lesssim 10^{-7}$-$10^{-5}$) than the suggested values of this work,  from double beta decay and SN1987a \cite{KamLAND-Zen:2012uen, Kachelriess:2000qc}.

Interpreting $\phi$ as the Majoron yields a prediction for the scale
of lepton-number ($L$) breaking and the associated heavy seesaw neutrinos. In singlet Majoron models, $L$ is spontaneously broken as described above Eq.\ (\ref{yukawa-lag}),
 generating Majorana masses for right-handed neutrinos through
\begin{equation}
\mathcal{L} \supset \frac{1}{2}\,h_{ij}\,\Phi\,\overline{N_i^c}N_j + \mathrm{h.c.}\,,
\end{equation}
where $N_i$ denote right-handed (sterile) neutrino fields, singlets under the
Standard Model gauge group. After symmetry breaking, the heavy neutrino masses are $M_{ij}=h_{ij} f/\sqrt{2}$, and the couplings of $\phi$ to neutrinos are fixed by $y_i=m_{\nu_i}/f$ in the mass basis.

The allowed range
$5\times10^{-16}\lesssim y \lesssim 10^{-12}$ implies
$f \sim 10^{11}$-$10^{14}\,{\rm GeV}$. For unsuppressed couplings $h_{ij} \sim \mathcal{O}(1)$, one gets
$M \sim 10^{2}$-$10^{5}\,{\rm TeV}$. The lower end of this range is consistent with low-scale leptogenesis, and with searches by beam dumps and colliders \cite{Akhmedov:1998qx,Drewes:2021nqr}.

\bigskip
\emph{\textbf{Conclusions}.} We have questioned the long-standing argument that Standard Model active neutrinos cannot constitute the dark matter of the Universe, and shown that this conclusion relies  on assumptions about their thermal origin and relativistic phase-space distribution. If  an enhanced population of neutrinos is produced nonthermally at late times, with sufficiently small momenta, active neutrinos can behave effectively as cold dark matter while remaining consistent with bounds from Big Bang nucleosynthesis, recombination, structure formation, and laboratory experiments. To our knowledge this is the first work that rescues active, Standard Model neutrinos as a dominant dark matter component.

We presented a minimal and predictive realization of this idea based on a light neutrinophilic scalar that never thermalizes with the plasma and decays into neutrinos after BBN.
The scalar can obtain the right relic abundance by the misalignment mechanism, which can satisfy constraints on the displacement of the scalar and the reheating temperature of the Universe. Other possible production mechanisms like decay of domain walls or inflatons are left for future study.

Post-recombination energy injection constraints on the neutrinos restricts the scalar mass to lie within a few percent of twice the neutrino mass, $m_\phi \lesssim 2.039\,m_\nu$. This fine-tuning might be ameliorated in 
chameleon or symmetron models, as described in the Supplementary Material \ref{sec:screening}. Requiring that scalar does not thermalize,
yet decays within 14\,Gyr, predicts Yukawa couplings in the range $5\times10^{-16}\lesssim y\lesssim10^{-12}$. In this regime, the C$\nu$B today is enhanced by two orders of magnitude relative to its standard cosmological value in the intergalactic medium.

Although fermionic phase-space bounds normally exclude sub-eV neutrinos as dark matter if they are free and collisionless during halo formation, we showed that this can be evaded. Either structure forms while the scalar field constitutes the dark matter, with neutrinos inheriting the dark matter density only at late times, or additional neutrino-sector interactions may allow neutrinos to form bosonic bound states at the relevant epochs.

The most striking consequence of our framework is an enhanced C$\nu$B on cosmological scales, which can be probed by upcoming efforts to detect relic neutrinos with a high-energy neutrino flux or gamma-ray flux arising from cosmic ray–neutrino scatterings. Our work demonstrates that active neutrinos remain a viable and testable dark matter candidate once the assumption of a thermal relic origin is relaxed.

\emph{\textbf{Acknowledgements}.} 
We are grateful to Jordi Salvado, Miguel Escudero and Gordan Krnjaic for useful feedback. The work of GH was supported by the Neutrino Theory Network Fellowship with contract number 726844, and by the U.S. Department of Energy under award number DE-SC0020262. This manuscript has been authored by FermiForward Discovery Group, LLC under Contract No. 89243024CSC000002 with the U.S. Department of Energy, Office of Science, Office of High Energy Physics.  JC and JSR are supported by the Natural Sciences and Engineering Research Council (NSERC) of Canada.

\bibliography{refs.bib}

\appendix

\section*{Supplementary Material}

\subsection{Isocurvature constraints}\label{sec:isocurvature_constraints}
The quantum fluctuations of the light scalar can generate isocurvature perturbations in the early Universe. For a pre-inflationary broken $U(1)$ with a homogeneous initial angle,
the field fluctuation is $\delta\phi \simeq H_I/(2\pi)$, implying an
isocurvature power spectrum
\begin{equation}
\mathcal P_S \simeq \left(\frac{\delta\phi}{\phi_0}\right)^2
\simeq \left(\frac{H_I}{\pi f \theta_0}\right)^2\,,
\end{equation}
where $\theta_0 \equiv \phi_0/f$ is the initial misalignment angle.
CMB data constrain the primordial CDM-isocurvature fraction to be at the
percent level, $\beta_{\rm iso}\equiv \mathcal P_S/(\mathcal P_{\mathcal R}+\mathcal P_S)
\lesssim 0.013$ \cite{Planck:2018jri}.
Using $\mathcal P_{\mathcal R}\equiv A_s \simeq 2.1\times10^{-9}$, this gives
\begin{equation}
H_I \;\lesssim\; \pi f \theta_0 \sqrt{\beta_{\rm iso} A_s}
\;\simeq\; 1.6\times10^{-5}\, f\,\theta_0\,
\left(\frac{\beta_{\rm iso}}{0.013}\right)^{1/2}\,.
\end{equation}

This isocurvature constraint is easily satisfied for values of the lepton number breaking scale $f$ consistent with the Majoron decaying in the right cosmological window (see next Section). For $f \sim 10^{11}-10^{14}$\,GeV, $H_I \lesssim 10^{6}-10^{9}$GeV, which is consistent with the value needed for the stochastic misalignment mechanism, of $H_I \sim 15$ GeV.

\subsection{Post recombination constraints on scalar decays into neutrinos}\label{sec:fradiation}

If neutrinos are produced relativistically in $\phi$ decays, they may reduce structure formation at small scales. We consider the conservative scenario in which
$\phi$ constitutes the dark matter after recombination and decays nonrelativistically as $\phi\to\nu\bar{\nu}$. CMB limits on post-recombination conversion of dark matter into invisible radiation imply that only a fraction $F_{\max}$ of the dark matter can become radiation-like between recombination and today
\cite{Poulin:2016nat,Bringmann:2018jpr},
\begin{equation}
F_{\max} \simeq 0.038\, .
\label{eq:fmax}
\end{equation}

We start with a simplified scenario, assuming a single (degenerate) neutrino species of mass $m_\nu$.
For a two-body decay at rest, the neutrino velocity is
\begin{equation}
v_\nu = \sqrt{1-\frac{4m_\nu^2}{m_\phi^2}}\,,
\end{equation}
corresponding to an equation of state $w_\nu\simeq v_\nu^2/3$. If a fraction $F=1$ of the dark matter decays after recombination, a conservative mapping onto the CMB bound gives
\begin{equation}
F_{\rm rad,eff}=v_\nu^2\lesssim F_{\max}\,.
\end{equation}
This implies
\begin{equation}
m_\phi \lesssim \frac{2m_\nu}{\sqrt{1-F_{\max}}}
\simeq 2.039\,m_\nu\,,
\end{equation} which is shown as a vertical purple line in Fig.~\ref{fig:parameter_space}.

In reality, $\phi$ decays into three neutrino mass eigenstates. In the Majoron realization with diagonal couplings $y_i=m_{\nu_i}/f$, the partial widths scale as
\begin{equation}
\Gamma_i \propto m_{\nu_i}^2\,v_i\,,
\qquad
v_i=\sqrt{1-\frac{4m_{\nu_i}^2}{m_\phi^2}}\, .
\end{equation}

We next improve on this in the more realistic case
with multiple decay channels. The effective radiation fraction is
\begin{equation}
F_{\rm rad,eff}
=
F\sum_i {\rm BR}_i\,v_i^2\,,
\qquad
{\rm BR}_i=\frac{\Gamma_i}{\Gamma_{\rm tot}}\, .
\label{eq:fradeff_multi}
\end{equation}
For fixed neutrino masses, $F_{\rm rad,eff}$ depends on the scalar mass
through the velocities $v_i$, and exhibits a tradeoff: choosing $m_\phi$ closer to threshold suppresses the velocities of the
heavier eigenstates but enhances their phase-space suppression (see Eq.\ (\ref{eq:decay_rate})), increasing the branching into lighter (relativistic) eigenstates; choosing $m_\phi$ larger reduces this branching but makes the dominant channels more relativistic. We then find the most permissive constraint by
minimizing $F_{\rm rad,eff}$ with respect to $m_\phi$. For normal ordering, $m_{\nu_3}\gg m_{\nu_2}>m_{\nu_1}$. Approximating $v_{1,2}\simeq 1$,
\begin{equation}
{\rm BR}_2 \simeq \frac{m_{\nu_2}^2}{m_{\nu_2}^2+m_{\nu_3}^2 v_3}
=\frac{r}{r+v_3},
\qquad
{\rm BR}_3 \simeq \frac{v_3}{r+v_3},
\end{equation}
with $r\equiv m_{\nu_2}^2/m_{\nu_3}^2\simeq 3\times 10^{-2}$.
Hence
\begin{equation}
F_{\rm rad,eff}(F=1)
=
{\rm BR}_2\,v_2^2+{\rm BR}_3\,v_3^2
\simeq
\frac{r+v_3^3}{r+v_3}\,,
\label{eq:fradeff_NO_approx}
\end{equation}
which incorporates both the branching-ratio and velocity effects. Minimizing this expression with respect to $v_3$ yields
\begin{equation}
F_{\rm rad,eff}^{\rm min}\simeq 0.16\,,
\end{equation}
well above the maximum allowed value $F_{\max}=0.038$.

Thus, in the post-recombination decay scenario with $F\simeq1$, normal ordering cannot satisfy the CMB bound, regardless of how close $m_\phi$ is taken to threshold. Normal mass ordering can nevertheless be viable if decays occur predominantly before recombination, if only a fraction $F<1$ of the dark matter decays at late times, or in models where the scalar coupling to neutrinos is more strongly suppressed than linearly with the eigenstate mass.

The situation is more favorable for inverted ordering, with $m_{\nu_1}\simeq m_{\nu_2}\gg m_{\nu_3}$.
Choosing $m_\phi\simeq2m_{\nu_1}\simeq2m_{\nu_2}$ yields two near-threshold
channels that have nearly the same decay width.
Defining
$\kappa\equiv m_{\nu_3}^2/(m_{\nu_1}^2+m_{\nu_2}^2)$, one finds
\begin{equation}
F_{\rm rad,eff}(F=1)
\simeq
\frac{\kappa+v_h^3}{\kappa+v_h}\,,
\end{equation}
where $v_h$ is the common velocity of the heavy eigenstates. For minimal inverted ordering ($m_{\nu_3}\to0$), $\kappa\to0$ and
$F_{\rm rad,eff}\to v_h^2$, so the post-recombination constraint reduces to the single-channel bound $m_\phi\lesssim2.039\,m_h$.
For small but nonzero $m_{\nu_3}$, minimizing $F_{\rm rad,eff}$ gives
$F_{\rm rad,eff}^{\rm min}\simeq1.9\,\kappa^{2/3}$, implying
\begin{equation}
\label{mnu3bound}
m_{\nu_3} \lesssim 3.7 \times 10^{-3} \mathrm{eV}\left(\frac{m_h}{0.05 \mathrm{eV}}\right)\,,
\end{equation}
if the decays occur predominantly after recombination. 

The inequality (\ref{mnu3bound}) is a mild requirement on the lightest neutrino eigenstate mass. We then conclude that inverted ordering naturally satisfies the post-recombination structure-formation constraint for scalar masses close to threshold, while
normal ordering requires earlier decays or a subdominant late-time conversion fraction. In Fig. \ref{fig:recombination_bound} we show the numerical effective radiation fraction as a function of the parent scalar mass, for both normal and inverted orderings, and confront it with the Planck limit from \cite{Poulin:2016nat,Bringmann:2018jpr}.

\begin{figure}[H]
    \centering
    \includegraphics[width=0.5\textwidth]{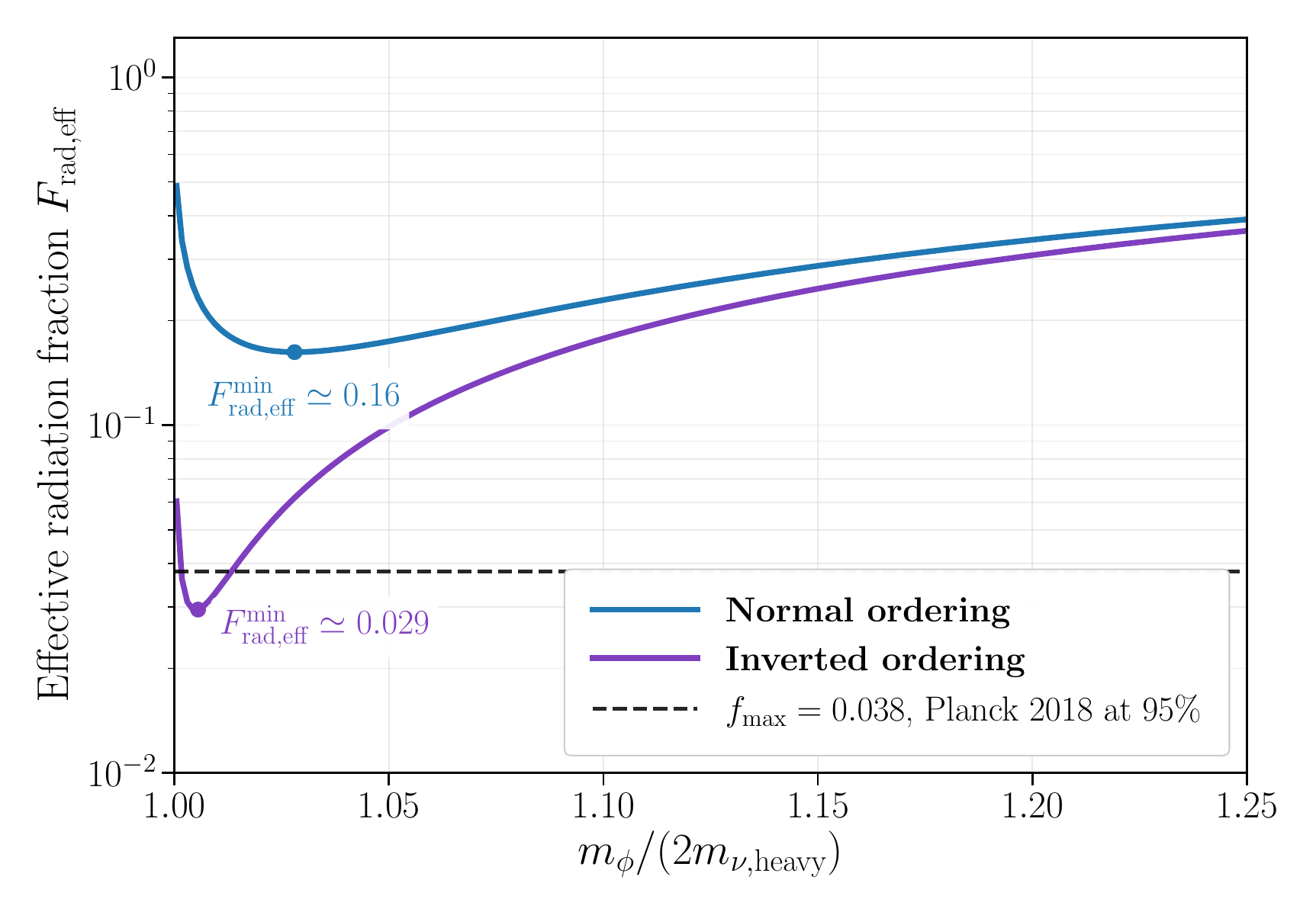} 
    \caption{Effective post-recombination radiation fraction from scalar decays into neutrinos as a function of scalar mass over twice the neutrino mass. Normal Ordering (NO) fails to satisfy the bound from \cite{Poulin:2016nat, Bringmann:2018jpr} for any mass, but inverted ordering (IO) does satisfy the bound.}
    \label{fig:recombination_bound}
\end{figure}
\subsection{Phase space constraints in small scales}\label{sec:pauli}

Independently of free-streaming and background cosmology, fermionic dark matter is subject to a phase-space constraint arising from the Pauli exclusion principle: the coarse-grained phase-space density in any virialized system cannot exceed the primordial fine-grained maximum. Applied to dwarf spheroidal galaxies, this Tremaine-Gunn bound implies that free, collisionless
fermions constituting the dominant halo dark matter must have masses
$m_{\rm DM}\gtrsim100\,\mathrm{eV}$ \cite{Tremaine:1979we,DiPaolo:2017geq,Alvey:2020xsk}, thereby excluding sub-eV active neutrinos as the dominant dark matter component if they are free and collisionless during halo formation.

The constraint follows directly from Fermi statistics. For fermions of mass $m_\nu$ in a gravitational potential characterized by a escape velocity $v_{\rm esc}$, the Pauli exclusion principle implies a maximum occupation of phase space $f(p)\le 2$, including spin. Approximating the distribution as filled up to a maximum (Fermi) momentum
$p_{F}\simeq m_\nu v_{\rm esc}$, the largest allowed neutrino number density is
\begin{equation}
n_{\nu}^{\rm max}
=
\frac{2}{(2\pi)^3}\int_0^{p_{F}}4\pi p^2\,dp
=
\frac{m_\nu^3\,v_{\rm esc}^3}{3\pi^2},
\end{equation}
corresponding to a maximal mass density
\begin{equation}
\rho_{\nu}^{\rm max}
=
m_\nu n_{\nu}^{\rm max}
=
\frac{m_\nu^4\,v_{\rm esc}^3}{3\pi^2}.
\label{eq:rho_max}
\end{equation}
This result depends only on gravity and Fermi statistics, and is independent of the neutrino production mechanism; it is the basis of the Tremaine-Gunn bound, which gives maximally optimistic estimates of neutrino clustering \cite{Bondarenko:2023ukx}.

In a scenario where the dark matter is initially composed of a finely-tuned scalar field $\phi$ that decays into neutrinos at late times, Pauli exclusion implies that decays can proceed only until the local neutrino density reaches $\rho_{\nu}^{\rm max}$. If the scalar mass is not sufficiently fine-tuned and the decayed neutrinos are not gravitationally bound, then decays may proceed, depleting the halo mass to unacceptable levels. Here we quantify these restrictions on various physical scales in a model-independent manner, and in the next section we describe a concrete model where corresponding environmentally dependent couplings could be realized. 

The maximal fraction of the dark matter density that can be converted into cold neutrinos in a given environment, by means of Pauli-exclusion principle is
\begin{equation}
F_{\rm Pauli}^{\rm max}
=
\frac{\rho_{\nu}^{\rm max}}{\rho_{\rm DM}}
=
\frac{m_\nu^4\,v_{\rm esc}^3}{3\pi^2\,\rho_{\rm DM}}\,,
\label{eq:Fmax}
\end{equation}
where $\rho_{\rm DM}$ is the (average) ambient dark matter density of the system. Numerically
\begin{equation}
F_{\rm Pauli}^{\rm max}
\simeq
1.6\times10^{-3}\,
\left(\frac{m_\nu}{0.1\,\mathrm{eV}}\right)^4
\left(\frac{v_{\rm esc}}{500\,\mathrm{km\,s^{-1}}}\right)^3
\left(\frac{\bar\rho_{\rm DM}}{\rho_{\rm DM}}\right).
\end{equation}
with $\bar\rho_{\rm DM}\equiv \Omega_{\rm DM}\rho_c \simeq 1.3\times10^{-6}\,{\rm GeV\,cm^{-3}}$. As a result, the Pauli constraint has very different implications depending on the physical scale. 

For the largest possible neutrino mass allowed by KATRIN \cite{KATRIN:2024cdt}, $m_{\nu}=0.4$ eV, in dwarf spheroidal galaxies, with $v_{\rm esc}\sim30$-$50\,\mathrm{km\,s^{-1}}$ and $\rho_{\rm DM}\sim0.1$-$1\,\mathrm{GeV\,cm^{-3}}$ \cite{Walker, Battaglia:2013wqa}, one finds
$F_{\rm Pauli}^{\rm max}\lesssim10^{-11}$-$10^{-10}$, implying that neutrinos cannot constitute an appreciable fraction of the halo dark matter.
In Milky-Way-like galaxies, with $v_{\rm esc}\sim500$-$600\,\mathrm{km\,s^{-1}}$ and $\rho_{\rm DM}\simeq0.1$-$1\,\mathrm{GeV\,cm^{-3}}$
\cite{Read:2014qva, Monari_2018}, one finds
$F_{\rm Pauli}^{\rm max}\sim10^{-7}$-$10^{-6}$, so that only a tiny fraction of the scalar can decay into neutrinos within the halo.%\js{I don't see how $10^{-6}$ is a percent-level fraction? Is this meant for clusters?}
In galaxy clusters, with $v_{\rm esc}\sim1500$--$3000\,\mathrm{km\,s^{-1}}$ and $\rho_{\rm DM}\simeq 10^{-3}$-$10^{-2}\,\mathrm{GeV\,cm^{-3}}$ \cite{Rines:2003tg}, the bound allows
$F_{\rm Pauli}^{\rm max}\sim 10^{-4}$-$10^{-2}$, while in the intergalactic medium, no Pauli constraint applies and neutrinos may constitute essentially all of the dark matter. The maximally allowed decayed dark matter fraction versus the escape velocity of the system is illustrated in Fig.\ \ref{fig:TG_neutrinos} for different values of $\rho_{\rm DM}$, and for two representative values of $m_{\nu}$ consistent with the laboratory bound from KATRIN ($m_{\nu} \lesssim 0.4$ eV) \cite{KATRIN:2024cdt}, and cosmological observations ($m_{\nu} \lesssim 0.1$ eV) \cite{Loureiro:2018pdz}.

This is a dynamically interesting and fairly novel physical picture: the scalar decays proceed until the neutrino phase space saturates, at which point further decays are Pauli-blocked. As a consequence, neutrinos cannot dominate in deep gravitational potentials, but remain viable as dark matter on large and diffuse scales.

Cosmological simulations and cosmic-web classifications indicate that only
$\sim 20$-$30\%$ of the total dark matter mass resides in galactic halos of
Milky-Way size and below, while $\sim 10$-$15\%$ is bound in clusters and the
remaining $\sim 50$-$70\%$ is distributed in diffuse structures outside
virialized galaxies \cite{Nuza:2014dua,Cautun:2014fwa, Haider:2015caa}. Therefore, even after accounting for Pauli blocking on
small scales, the majority of the cosmological dark matter (potentially at the level of 50-70$\%$), can be made of cold active neutrinos. This leads to a strongly enhanced C$\nu$B on cosmological scales.

There are two qualitatively distinct ways to avoid the traditional Tremaine-Gunn constraint. First, structure formation may proceed while the scalar field $\phi$ constitutes the dark
matter, with its decay into neutrinos occurring only after the formation of the relevant gravitational potentials. Since the Tremaine-Gunn argument assumes that halos are composed of free fermions from the outset, it does not apply if halos are initially formed by $\phi$ and neutrinos inherit the dark matter density only at late times. The Pauli exclusion principle still prevents too many neutrinos from being gravitationally bound in a dwarf spheroidal galaxy, independently of the history of structure formation. Requiring decay after dwarf-halo assembly
($z\sim10$-$20$ \cite{Barkana:2000fd}) or after Milky-Way-like galaxy assembly
($z\sim1$-$3$ \cite{Wechsler:2001cs}) motivates the lifetime contours shown in Fig.~\ref{fig:parameter_space}.
Second, additional neutrino-sector interactions, such as strong self-interactions mediated by a very light or massless scalar, may allow neutrinos to behave as a nonfree fermionic fluid or to form bound states, thereby invalidating the assumptions underlying the Tremaine-Gunn bound \cite{Smirnov:2022sfo,Ferrer:1999ad, Esteban:2021ozz}.

\begin{figure}[H]
    \centering
    \includegraphics[width=0.5\textwidth]{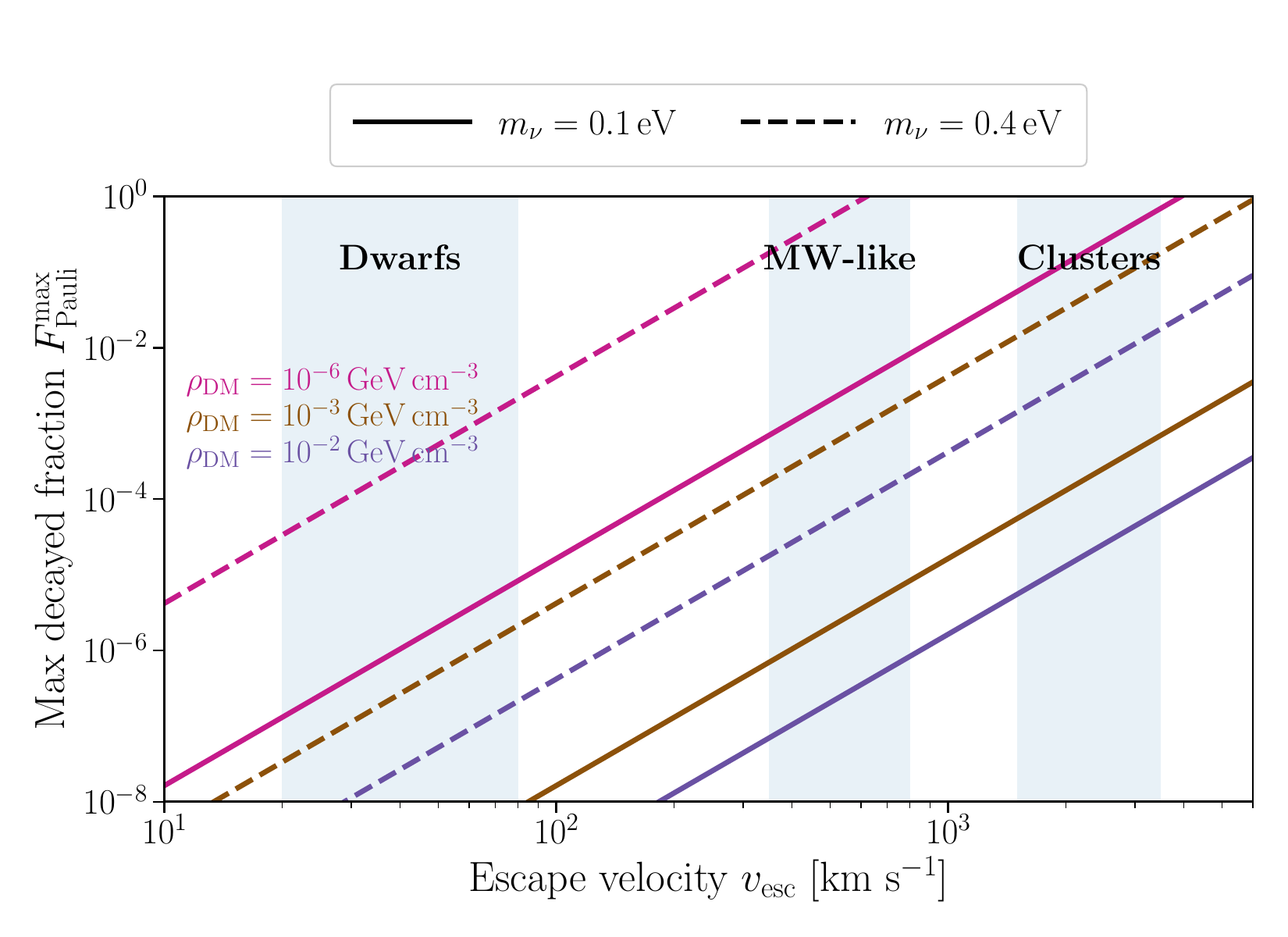} 
    \caption{Maximally allowed fraction of dark matter that can decay into free active neutrinos without violating the Pauli exclusion principle (Tremaine–Gunn bound), as a function of the escape velocity of the system. Curves correspond to different neutrino masses, while shaded bands indicate characteristic velocities of dwarf galaxies, Milky-Way–like halos, and galaxy clusters. Neutrino dark matter is strongly suppressed in galactic halos but allowed on large, diffuse cosmological scales.}
    \label{fig:TG_neutrinos}
\end{figure}

A further, independent consideration concerns whether the neutrinos produced in $\phi \rightarrow \nu \bar{\nu}$ decays behave as cold matter on small scales. If the decayed neutrinos constitute the bulk of the dark matter on such scales, then they must be sufficiently slow by the time small-scale structures form.

For a two-body decay at rest, the neutrino momentum at production is
\begin{equation}
p_\star = \frac{m_\phi}{2}\,\beta,
\qquad
\beta = \sqrt{1-\frac{4m_\nu^2}{m_\phi^2}},
\end{equation}
corresponding to a velocity
\begin{equation}
v_\nu^{\rm prod}
= \frac{p_\star}{\sqrt{p_\star^2+m_\nu^2}} \, .
\end{equation}
If the decay occurs at redshift $z_{\rm dec}$, cosmological expansion reduces
the momentum according to
\begin{equation}
p(z) = p_\star \, \frac{1+z}{1+z_{\rm dec}} \, .
\end{equation}
Thus, the velocity relevant for gravitational binding is not the production velocity but the velocity at the epoch when small halos assemble. We conservatively evaluate this at a characteristic dwarf-galaxy formation redshift $z_f \simeq 20$, requiring
\begin{equation}
v_\nu(z_f) \;\lesssim\; v_{\rm esc}^{\rm dwarf},
\end{equation}
where $v_{\rm esc}^{\rm dwarf} \lesssim 80~\mathrm{km\,s^{-1}}$ is a
representative escape velocity for dwarf halos.

This condition links the Yukawa coupling to the scalar mass. Earlier decays lead to larger redshift cooling and hence smaller velocities at $z_f$. We show the allowed parameter space on the yukawa coupling versus $\delta \equiv (m_\phi - 2 m_\nu)/2 m_\nu$ in Fig. \ref{fig:dwarf-coldness}.

\begin{figure}[H]
    \centering
    \includegraphics[width=0.5\textwidth]{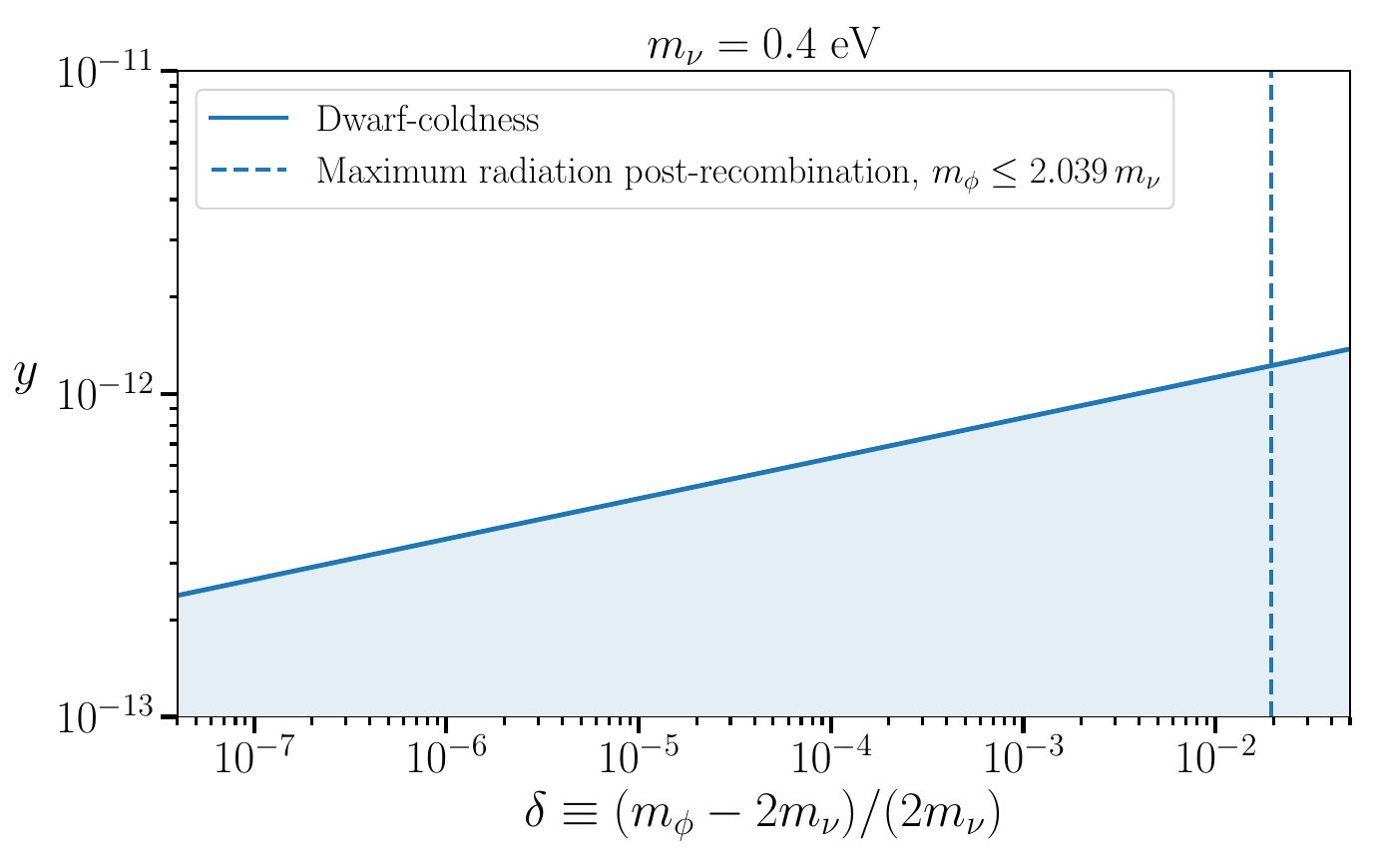} 
    \caption{Dwarf-coldness constraint on late $\phi \to \nu\bar{\nu}$ decays for $m_\nu = 0.4~\mathrm{eV}$. The blue curve shows the boundary obtained by requiring the neutrino velocity at dwarf-galaxy formation ($z_f \simeq 20$) to satisfy $v_\nu(z_f) = v_{\rm esc}^{\rm dwarf}=80~\mathrm{km\,s^{-1}}$. Parameter space above the curve correspond to earlier decays and sufficient cosmological redshifting of the decay products, such that neutrinos behave as cold matter on dwarf scales, while the region below corresponds to too warm neutrinos. The vertical dashed line shows the independent post-recombination constraint $m_\phi \le 2.039\,m_\nu$.} %\js{Maybe we should shade the region that is ruled out (below the solid curve) to be consistent with Fig. 1}
    \label{fig:dwarf-coldness}
\end{figure}

\begin{figure*}[t]
    \centering
    \includegraphics[width=0.7\textwidth]{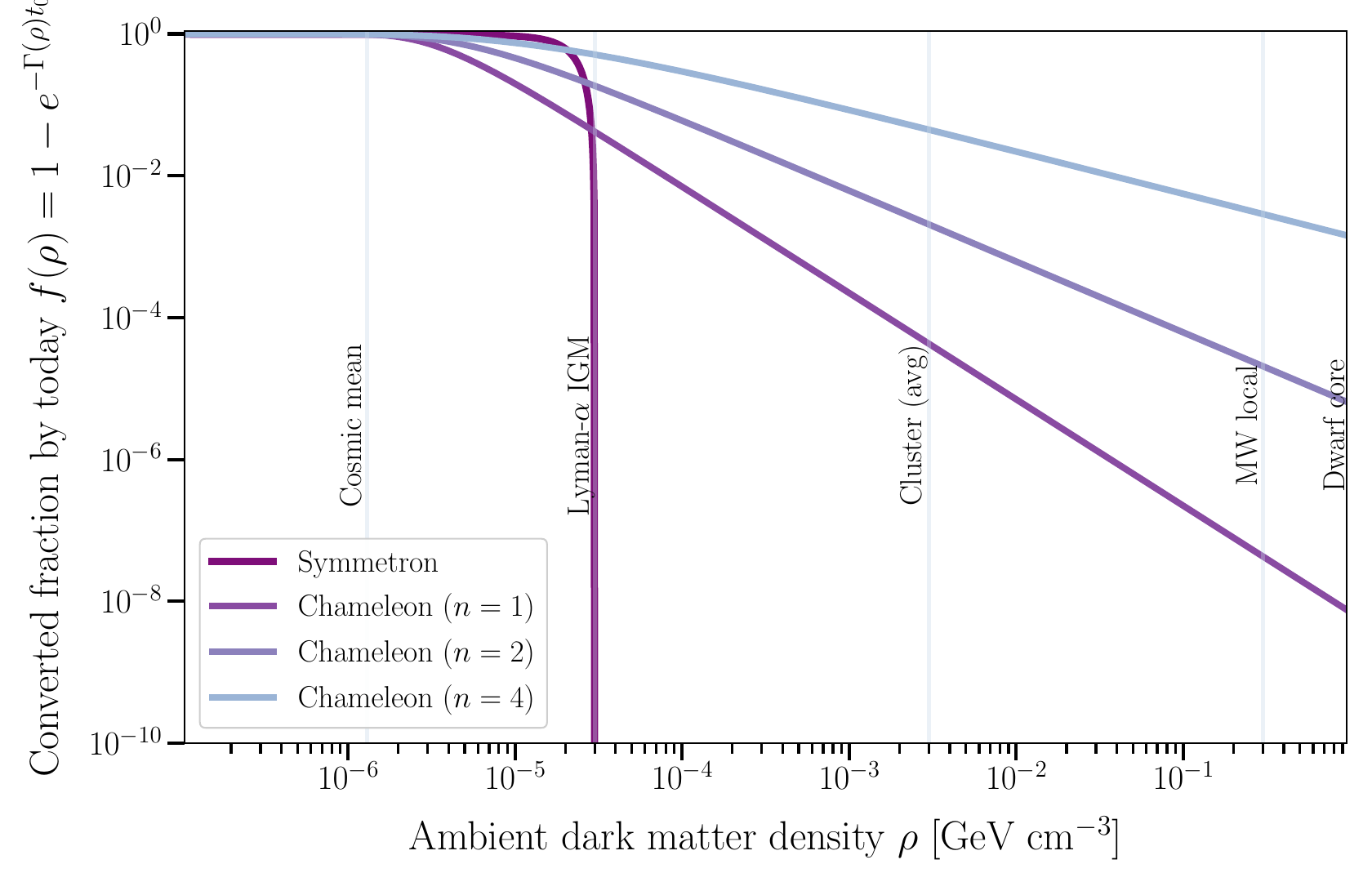} 
    \caption{
Converted fraction $f(\rho)=1-\exp[-\Gamma_{\rm tot}(\rho)t_0]$ of scalar dark matter into neutrinos by today as a function of ambient density, for a symmetron-gated mass $m_\phi(\rho)$ (solid) and chameleon-gated masses (colored) with indices $n=\{1,2,4\}$.  The Yukawa scale is fixed in each case by requiring near-complete conversion on diffuse scales, $f(\bar\rho)=0.99$ at the cosmic mean density $\bar\rho\simeq1.3\times10^{-6}\,\mathrm{GeV\,cm^{-3}}$.  Vertical lines indicate representative densities for different environments.  Screening dynamically suppresses decays in dense halos while allowing efficient conversion in diffuse regions, yielding a mixed scalar-neutrino dark matter composition that depends on physical scale and relaxing the need for a globally fine-tuned scalar mass near $m_\phi\simeq 2m_\nu$.
}   \label{fig:chameleon}
\end{figure*}

\subsection{Chameleons and Symmetrons to ameliorate the scalar mass fine-tuning}
\label{sec:screening}
The bounds discussed above motivate scenarios in which the late-time conversion
$\phi\to \nu\bar{\nu}$ is efficient on diffuse (cosmological) scales, while being
strongly suppressed inside dense environments such as galaxies and dwarfs.
This can be achieved if the scalar mass is not a universal constant, but instead
depends on an environmental screening field $\chi$ whose expectation value varies
with the ambient matter density. In such a setup, the decay can be kinematically
``gated'': in low-density regions the detuning from threshold is relatively large
and the decay is fast, whereas in high-density regions the scalar mass is
dynamically driven close to $2m_\nu$, suppressing the phase space and effectively
stabilizing $\phi$ inside halos.

This mechanism trades a global fine-tuning of $m_\phi$ for an environmentally
induced relation $m_\phi(\rho)$. We describe the coupling of $\phi$ to $\chi$
through
\begin{equation}
\mathcal{L} \supset -\frac{1}{2}\, m_\phi^2(\chi)\,\phi^2\,,
\qquad
m_\phi^2(\chi) = m_0^2 + g^2\chi^2\,,
\label{eq:mphi_screening}
\end{equation}
where $m_0\simeq 2m_\nu$ is the high-density value and $g$ sets the strength of the
cross-coupling. Such a dependence can arise, for example, in Majoron-like
realizations where an explicit $U(1)$-breaking operator depends on $\chi$, as
discussed in the main text.
Given $m_\phi(\rho)\equiv m_\phi(\langle\chi(\rho)\rangle)$, the two-body decay width
into a neutrino mass eigenstate $\nu_i$ is
\begin{equation}
\Gamma_i(\rho)=\frac{y_i^2\,m_\phi(\rho)}{8\pi}\,v_i(\rho)^3\,,
\quad
v_i(\rho)=\sqrt{1-\frac{4m_{\nu_i}^2}{m_\phi(\rho)^2}}\,,
\label{eq:Gamma_i_rho_app}
\end{equation}
so that near threshold the decay rate inherits a strong dependence on the
detuning. Defining
\begin{equation}
\delta_i(\rho)\equiv \frac{m_\phi(\rho)-2m_{\nu_i}}{2m_{\nu_i}}\,,
\end{equation}
one has $v_i^2\simeq 2\delta_i$ for $\delta_i\ll 1$, implying the characteristic
scaling
\begin{equation}
\Gamma_i(\rho)\propto y_i^2\,m_{\nu_i}\,\delta_i(\rho)^{3/2}\,.
\label{eq:Gamma_scaling_delta}
\end{equation}
Thus, driving $\delta_i(\rho)$ small in halos leads to a large suppression of
$\Gamma_i(\rho)$, even if $\delta_i(\rho)$ is much larger in diffuse regions.
For a given environment, the fraction converted by today is
\begin{equation}
f(\rho)=1-\exp\!\left[-\Gamma_{\rm tot}(\rho)\,t_0\right]\,,
\qquad
\Gamma_{\rm tot}(\rho)=\sum_i \Gamma_i(\rho)\,,
\label{eq:f_rho}
\end{equation}
with $t_0$ the age of the Universe. In our phenomenological exploration we fix
the Yukawa scale by requiring near-complete conversion at the cosmic mean
density, $f(\bar{\rho})=0.99$, where
$\bar{\rho}\simeq\Omega_{\rm DM}\rho_c\simeq 1.3\times 10^{-6}\,
{\rm GeV\,cm^{-3}}$. This fixes the overall normalization of the $y_i$ once
$m_\phi(\bar{\rho})$ is specified. With this choice, the same model predicts
$f(\rho)\ll 1$ in high-density environments if $m_\phi(\rho)$ is driven close
enough to threshold.

There are two known screening mechanisms that can generate the required
behavior of $\langle\chi(\rho)\rangle$. We discuss both below and provide
quantitative estimates of the cross-coupling $g$ needed in each case.

\medskip
\noindent\textbf{Symmetron.}
In symmetron models~\cite{Hinterbichler:2010es}, the effective potential in the
presence of ambient density $\rho$ takes the form
\begin{equation}
V_{\rm eff}(\chi,\rho)
= -\frac{1}{2}\mu^2\chi^2 + \frac{1}{4}\lambda_\chi\chi^4
  + \frac{\rho}{2M^2}\chi^2\,,
\label{eq:symmetron_Veff_app}
\end{equation}
leading to
\bea
\langle\chi(\rho)\rangle &=&
\begin{cases}
  0\,, & \rho > \rho_\star\,,\\[3pt]
  \chi_0\sqrt{1-\rho/\rho_\star}\,, & \rho < \rho_\star\,,
\end{cases}\nn\\
\rho_\star &\equiv& \mu^2 M^2\,,
\qquad
\chi_0 \equiv \mu/\sqrt{\lambda_\chi}\,.
\label{eq:symmetron_vev_app}
\eea
This interpolates between $m_\phi(\rho) = m_0$ at high density and a larger mass
at low density. The gating mechanism requires the critical density to lie between
intergalactic and halo scales,
$\bar{\rho} \lesssim \rho_\star \lesssim \rho_{\rm halo}$.
Since the astrophysical density range spans
$\bar{\rho}\sim 10^{-6}\,{\rm GeV\,cm^{-3}}$ to
$\rho_{\rm halo}\sim 0.3\,{\rm GeV\,cm^{-3}}$,
there is a wide window of roughly six orders of magnitude in which $\rho_\star$
may lie.
For $\lambda_\chi \sim \mathcal{O}(1)$, the VEV is
$\chi_0 = \mu/\sqrt{\lambda_\chi} \sim \mu$, and the constraint
$\rho_\star = \mu^2 M^2 \sim 10^{-2}\,{\rm GeV\,cm^{-3}}$
determines $M$ in terms of $\mu$; for instance, $\mu \sim \text{meV}$
gives $M \sim 0.3\,$eV.
When $\rho_{\rm halo} > \rho_\star$, the VEV vanishes exactly in halos,
$\langle\chi(\rho_{\rm halo})\rangle = 0$, so that $m_\phi = m_0$ and the
decay is completely shut off ; the suppression is formally infinite.
Matching $f(\bar{\rho}) = 0.99$ then requires a cross-coupling
$g \sim 10^{-4}$--$10^{-7}$ depending on the Yukawa coupling $y$, with
radiative corrections
$\delta m_\phi^2/m_0^2 \sim g^2\mu^2/(8\pi^2 m_0^2) \sim 10^{-13}$,
well under control.
\medskip

\noindent\textbf{Chameleon.}
In chameleon models~\cite{Khoury:2003rn,Khoury:2003aq}, the screening field is
governed by an inverse-power potential,
\begin{equation}
V(\chi) = \frac{\Lambda^{4+n}}{\chi^n}\,,
\end{equation}
and couples to matter density through
\begin{equation}
\mathcal{L} \supset \frac{\chi}{M}\,\rho\,.
\end{equation}
The effective potential
\begin{equation}
V_{\rm eff}(\chi,\rho)
= \frac{\Lambda^{4+n}}{\chi^n} + \frac{\rho}{M}\,\chi
\end{equation}
has a density-dependent minimum determined by $dV_{\rm eff}/d\chi = 0$, giving
\begin{equation}
\langle\chi(\rho)\rangle
= \left(\frac{n\,M\,\Lambda^{4+n}}{\rho}\right)^{\!1/(n+1)}.
\label{eq:chameleon_vev}
\end{equation}
Thus the field value decreases with ambient density as
$\langle\chi(\rho)\rangle \propto \rho^{-1/(n+1)}$,
and the index $n$ controls how rapidly screening turns on. In particular,
\begin{equation}
m_\phi(\rho) - 2m_\nu \propto \rho^{-2/(n+1)}\,,
\end{equation}
so the decay rate inherits a strong density dependence.

Unlike the symmetron, the chameleon provides a smooth power-law suppression across
all densities with no need to tune a critical density into the astrophysical
window. For $n = 4$, $\Lambda \simeq 2.4\times 10^{-3}\,$eV (the dark-energy
scale), and $M = M_{\rm Pl}$, the field values at the cosmic mean and at a typical
halo density ($\rho_h \simeq 0.3\,{\rm GeV\,cm^{-3}}$) are
$\langle\chi(\bar{\rho})\rangle \sim 5\times 10^{3}\,$eV and
$\langle\chi(\rho_h)\rangle \sim 5\times 10^{2}\,$eV, respectively.
Matching $f(\bar{\rho}) = 0.99$ requires a cross-coupling
$g \sim 5\times 10^{-8}$, which yields a detuning
$\delta(\bar{\rho}) \sim 10^{-6}$ in diffuse regions and
$\delta(\rho_h) \sim 10^{-8}$ inside halos; a suppression of the decay rate by a
factor of $\sim\!140$. The one-loop radiative correction to the scalar mass from
the $g^2\phi^2\chi^2$ portal is of order
$\delta m_\phi^2 \sim g^2 m_\chi^2/(16\pi^2) \ll m_0^2$, so the mechanism is
radiatively stable. While the suppression is finite (unlike the symmetron),
the chameleon parameters $\Lambda$ and $M$ are independently motivated by
cosmology and gravity, making this realization more predictive.

In diffuse environments, $f(\rho)$ can be close to unity, producing a large
neutrino dark matter fraction and an enhanced cosmological neutrino density,
while on smaller scales the scalar remains the bound dark matter and the decayed
fraction is small because the product neutrinos can escape the gravitational
system. Figure~\ref{fig:chameleon} illustrates this behavior using representative
symmetron and chameleon density dependences for $n = \{1,2,4\}$.

\subsection{Experimental constraints and detection prospects on the C$\nu$B}\label{sec:testability_appendix}

Detecting the C$\nu$B directly is challenging, despite having an enhancement in the number density relative to $\Lambda$CDM, as proposed in the main text. A longstanding proposal is the capture on a beta-decaying nucleus, which may be performed with the KATRIN experiment or the future PTOLEMY experiment \cite{Weinberg:1962zza,KATRIN:2022kkv, PTOLEMY:2019hkd}. KATRIN currently constraints a local overdensity at the level of $\sim 10^{11}$ times the expected cosmological average in $\Lambda$CDM, and while PTOLEMY is expected to improve this limit by orders of magnitude, an actual detection may be harnessed by the quantum broadening of the beta-decayed electrons at the tail of the distribution, which may be larger than the neutrino mass \cite{Cheipesh:2021fmg,Nussinov:2021zrj,PTOLEMY:2022ldz}. Given the Tremaine-Gunn bound, the enhancement on the C$\nu$B within the Milky Way from the kinematically-allowed decays from the scalar would be of order $\sim 1\%$, much smaller than other possible enhancements, for instance due to gravitational clustering \cite{Ringwald:2004np}. 

Another promising proposal relies on cosmic-ray scatterings off the C$\nu$B, boosting it to (ultra)-high energies, where a detection with neutrino telescopes may be possible \cite{Ciscar-Monsalvatje:2024tvm,DeMarchi:2024zer, Herrera:2024upj,Zhang:2025rqh,Herrera:2026pzj}. The nondetection of ultra-high energy neutrinos at IceCube allows us to exclude overdensities of $\sim 10^{3}-10^{5}$ times the cosmological average, with uncertainties dominated by the cosmic ray flux evolution with redshift, which may follow the star formation fate \cite{Hopkins:2006bw} or the quasar evolution function \cite{Hook:2002st}. Comparable bounds are obtained from the non-observation of an excess gamma-ray flux arising from cosmic ray-relic neutrino scatterings in Fermi-LAT \cite{Herrera:2026bie}.

We underscore that our proposal may feature mixed neutrino-scalar populations behaving as dark matter in certain environments, like dwarf galaxies or Milky Way-like galaxies, where Pauli exclusion principle only permits a small fraction of the scalar population to decay into neutrinos. In other diffuse structures, such as clusters, superclusters and the intergalactic medium, active neutrinos may be the sole dark matter component. This may lead to predictable anisotropies in the cosmic-ray boosted active neutrino component.

\end{document}